\documentclass[twocolumn,showpacs,aps,prb,superscriptaddress,floatfix,amsmath,amssymb]{revtex4}
\usepackage{graphicx}% Include figure files
\usepackage{dcolumn}% Align table columns on decimal point
\usepackage{bm}% bold math
\usepackage{subfigure}

\begin{document}

\title{Magnetic order and ice rules in the multiferroic spinel $\mathrm{FeV_{2}O_{4}}$.}

\author{G. J. MacDougall}
\email{macdougallgj@ornl.gov}
\affiliation{Quantum Condensed Matter Division, Oak Ridge National Laboratory, Oak Ridge, Tennessee, 37831, USA}

\author{V. O. Garlea}
\affiliation{Quantum Condensed Matter Division, Oak Ridge National Laboratory, Oak Ridge, Tennessee, 37831, USA}

\author{A. A. Aczel}
\affiliation{Quantum Condensed Matter Division, Oak Ridge National Laboratory, Oak Ridge, Tennessee, 37831, USA}

\author{H. D. Zhou}
\affiliation{Department of Physics and Astronomy, University of Tennessee, Knoxville, TN, 37996, USA}
\affiliation{National High Magnetic Field Laboratory, Tallahassee, Florida 32310, USA}

\author{S. E. Nagler}
\affiliation{Quantum Condensed Matter Division, Oak Ridge National Laboratory, Oak Ridge, Tennessee, 37831, USA}
\affiliation{CIRE, University of Tennessee, Knoxville, TN, USA, 37996}

\date{\today}
\begin{abstract}

We present a neutron diffraction study of $FeV_{2}O_{4}$, which is rare in exhibiting spin and orbital degrees of freedom on both cation sublattices of the spinel structure. Our data confirm the existence of three structural phase transitions previously identified with x-ray powder diffraction, and reveal that the lower two transitions are associated with sequential collinear and canted ferrimagnetic transitions involving both cation sites. Through consideration of local crystal and spin symmetry, we further conclude that $Fe^{2+}$ cations are ferro-orbitally ordered below 135K and $V^{3+}$ orbitals order at 60K in accordance with predictions for vanadium spinels with large trigonal distortions and strong spin-orbit coupling. Intriguingly, the direction of ordered vanadium spins at low temperature obey `ice rules' more commonly associated with the frustrated rare-earth pyrochlore systems.
\end{abstract}

%PACS numbers {neutron diffraction for structures, spin arrangements in magnetically ordered materials, orbital charge or other orders, magnetic oxides

\maketitle

Magnetic insulators with orbital degrees-of-freedom are of great interest to both the condensed matter and applied physics communities. Transition-metal spinels are a particular example where orbital degeneracy and the concomitant orbital ordering have been shown to profoundly affect both the crystal structure and magnetic exchange interactions\cite{radaelli04,lee10}. In recent years, the material $\mathrm{FeV_{2}O_{4}}$ has emerged as an important model system to explore this physics. A rare example of a spinel with two orbitally-active cation sites,  $\mathrm{FeV_{2}O_{4}}$ is a multiferroic, containing both ferroelectric\cite{zhang12} and ferrimagnetic moments\cite{rogers63,tanaka66,katsufuji08,zhang12} at low temperatures. Moreover, separate experiments have demonstrated that the applied magnetic fields can both suppress the ferroelectric moments\cite{katsufuji08} and induce structural detwinning\cite{katsufuji08}. Previous x-ray diffraction measurements\cite{katsufuji08} mapped out a sequence of structural phase transitions, however despite the obvious importance of magneto-elastic and magneto-electric coupling, magnetic structures have remained unknown. In this Letter, we present new neutron scattering data definitively solving the magnetic structures as a function of temperature. The results establish that $\mathrm{FeV_{2}O_{4}}$ contains unexpected and very interesting parallels to the well-known "spin-ice" problem of frustrated magnetism\cite{bramwell01_2}, realized at low temperatures in selected rare-earth containing pyrochlore materials.

The spinel vanadates, $\mathrm{AV_{2}O_{4}}$, with divalent A-site cations have been central to the study of orbital order and the role of orbital degrees-of-freedom in geometrically frustrated antiferromagnets. In these systems, the pyrochlore sublattice of the spinel structure is occupied by $V^{3+}$ (3$d^{2}$) cations in an octahedral oxygen environment, giving spin S=1 and an orbital triplet degree-of-freedom. Materials such as $\mathrm{ZnV_{2}O_{4}}$\cite{ueda97,reehuis03,lee04}, $\mathrm{CdV_{2}O_{4}}$\cite{nishiguchi02} and $\mathrm{MgV_{2}O_{4}}$\cite{plumier63,mamiya97,wheeler10}, where the A-site is magnetically neutral, are typified by two transitions: an orbital ordering transition which lowers crystal symmetry at an upper temperature, followed by a second magnetic transition to a state with a complex pattern of ordered spins. In $\mathrm{MnV_{2}O_{4}}$, with spin-only $Mn^{2+}$ on the A-site, the system first orders magnetically to a collinear spin state involving both cation sites, only to develop a non-collinear structure below a second transition at a lower temperature\cite{adachi05,suzuki07_2,garlea08,chung08}. Despite being electronically and structurally similar at high temperatures, the low temperature properties of these materials are strikingly different, as the relative importance of spin-orbit coupling, tetragonal and trigonal crystal fields vary in different systems.
In each of these materials, debate continues and centers on understanding the nature of the orbital order inferred from measured crystal symmetry and magnetic structure\cite{tsunetsugu03,tchernyshyov04,perkins07,pardo08,maitra07,sarkar09,chern10}. A largely separate line of research has concentrated on the role of orbitals on the A-site sublattice, for example in $\mathrm{FeCr_{2}O_{4}}$\cite{shirane64,goodenough64,tsuda10,bordacs09} or $\mathrm{FeSc_{2}S_{4}}$\cite{krimmel05,chen09}; however, very few materials have been identified with orbitally active ions on \textit{both} cation sites of the spinel structure.

In this regard, $\mathrm{FeV_{2}O_{4}}$ is an extraordinary material. In addition to $V^{3+}$ on the pyrochlore sublattice, this system contains tetrahedrally-coordinated $Fe^{2+}$ (3$d^{6}$) cations on the A-site of the spinel structure with high-spin S=2 and two-fold orbital degeneracy. As might be expected, further interactions result in additional complex and interesting behavior. Specific heat measurements have revealed three phase transitions in this material with $T_c$ = 138, 111 and 56 K\cite{zhang12}. Powder x-ray scattering\cite{katsufuji08} has associated these temperatures with a cascade of structural transitions, from cubic to high-temperature tetragonal (HTT) to face-centered orthorhombic (FCO) to low-temperature tetragonal (LTT) with decreasing temperature. Both M$\ddot{o}$ssbauer\cite{tanaka66} and magnetization\cite{katsufuji08,zhang12} measurements indicate that at least one of these transitions (HTT$\rightarrow$FCO) additionally involves the ordering of magnetic degrees-of-freedom. A fourth structural phase transition to low-temperature orthorhombic (LTO) symmetry has been reported in one single crystalline study\cite{katsufuji08}, and at least one AC-susceptibility study has identified a glass-like transition at 85 K\cite{nishihara10}. Despite these advances and the implied coupling of electric, lattice and magnetic degrees-of-freedom, very little is known about the local spin and orbital structure in the distinct phases of $\mathrm{FeV_{2}O_{4}}$.

In this Letter, we present results of a neutron powder diffraction (NPD) study of $\mathrm{FeV_{2}O_{4}}$ at temperatures below 200 K. We confirm the cubic-HTT-FCO-LTT sequence of structural phase transitions reported by powder x-ray scattering\cite{katsufuji08}, but see no evidence for the purported fourth transition to a low-temperature orthorhombic phase\cite{katsufuji08}. Hitherto unreported, we discuss the evolution of the crystal field environment of $V^{3+}$ cations, and in particular present evidence for a large trigonal distortion of the octahedral oxygen cage about these sites. We additionally report the existence of two separate magnetic ordering transitions: a collinear spin-ordering coincident with the HTT-FCO transition and a tilting of $V^{3+}$ moments to point along the $<$111$>$ crystal directions when symmetry changes from FCO to LTT. We discuss the implications of the current observations for the orbital order, and in particular argue that this material is well-described by models which assume large trigonal distortion and strong spin-orbit coupling.

Polycrystalline samples of $\mathrm{FeV_{2}O_{4}}$ were prepared by solid state reaction, and NPD measurements were performed using the HB2A diffractometer at the High Flux Isotope Reactor\cite{HB2a}.  Further elastic neutron scattering data was taken on a crystal grown via the float zone method using the CG-4C Cold Neutron Triple-Axis Spectrometer of HFIR. Details about sample preparation and experimental method can be found in the Supplementary Material, along with characterization data on both powder and single-crystal samples.

\begin{figure}[t]
\begin{center}
\includegraphics[width=\columnwidth]{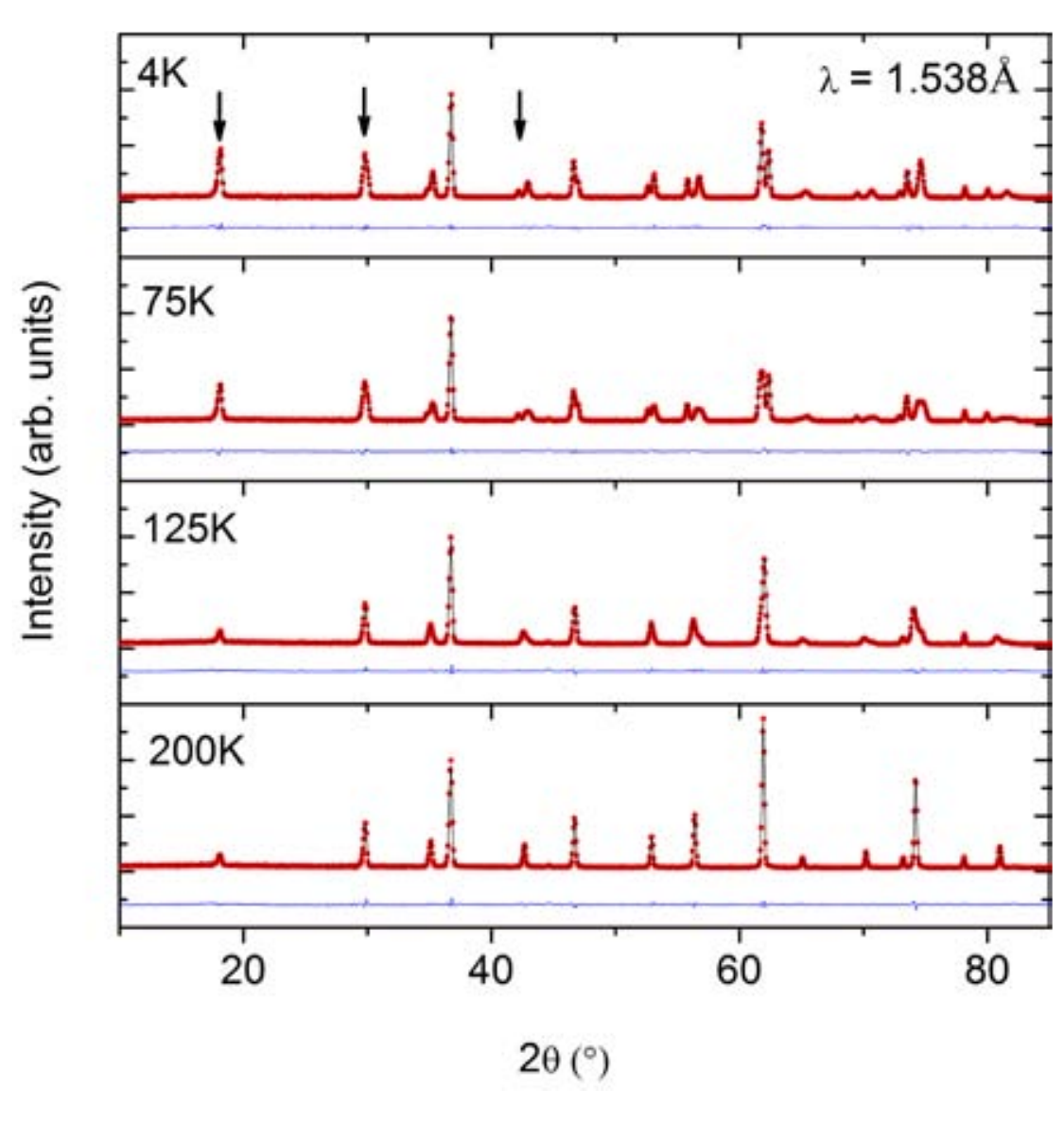}
\caption{Plots of raw NPD data (circles) measured at T = 200 K, 125 K, 75 K and 4 K for 2$\theta \le$ 85$^\circ$. Solid lines are results of Rietveld refinements described in the main text. Differences between observed and calculated intensities are shown directly below the respective patterns. See the main text and Figure~\ref{fig:evolution} for more detail.}\label{fig:raw}
\end{center}
\end{figure}

\begin{figure}[tbh]
\begin{center}
\includegraphics[width=\columnwidth]{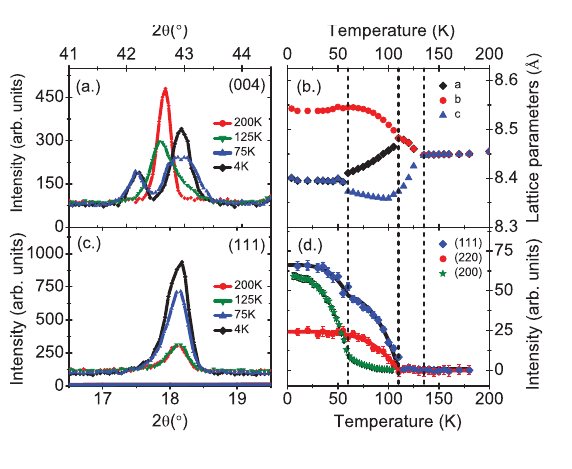}
\caption{\label{fig:evolution} \textbf{(a)} An expanded view of the NPD data around the cubic (004) Bragg position for each of the patterns in Fig.~\ref{fig:raw}. \textbf{(b)} Plot of the cubic lattice parameters extracted from NPD data. \textbf{(c)} An expanded view around the cubic (111) Bragg position for the same temperatures as (a). \textbf{(d)} A comparison of the temperature evolution of NPD (111) and (220) Bragg peak intensities, above values their values at 180 K. Overlaid is the scaled intensity of the (200) Bragg peak, as measured in a single crystal sample. }
\end{center}
\end{figure}

The results of our high-resolution neutron diffraction measurements at 200 K, 125 K, 75 K and 4 K are shown in Fig.~\ref{fig:raw} for 2$\theta \le$ 85$^\circ$. The highest temperature data confirm the spinel structure at these temperatures and the near phase purity. Refinement implies that a fraction of the vanadium sites is occupied by iron cations, giving chemical formula $\mathrm{Fe[Fe_{0.058}V_{0.942}]_2 O_4}$. This is similar to other samples in the literature\cite{katsufuji08,zhang12}. There is also a small temperature-independent impurity phase, possibly associated with an oxide of vanadium. The temperatures of remaining high-resolution patterns were chosen to be well within each of the five phases identified by Katsufuji \textit{et al.} with single-crystal x-ray diffraction\cite{katsufuji08}. Contrary to those results however, no discernable difference was seen between the patterns at 50 K and 4K, and so only the 4K pattern is plotted here. The four unique patterns were well-described by the space groups $Fd\bar{3}m$, $I4{1}/amd$, $Fddd$ and $I4{1}/amd$ with decreasing temperature, consistent with previous analysis and reflecting the cubic-HTT-FCO-LTT sequence of structural phase transitions\cite{katsufuji08}. Patterns to $2\theta=154^\circ$ and full refinement information are given as Supplementary Material, along with a more in-depth discussion of the absent LTT-LTO phase transition.

The sequence of observed structural transitions is illustrated more clearly in Fig.~\ref{fig:evolution}(a) and (b).  Fig.~\ref{fig:evolution}(a) shows the successive splittings and recombination of the cubic (004) peak for each of the four patterns. Fig.~\ref{fig:evolution}(b) shows a plot of the lattice parameters extracted from lower-resolution NPD data assuming the crystal symmetries determined above. (For the sake of comparison, all lattice parameters plotted here are those for the cubic unit cell.) Transition temperatures are identified as 135 K, 110 K and 60 K, within an uncertainty of 5 K. These estimates are similar to the values inferred from heat capacity in recent measurements of powders by Zhang \textit{et al.}\cite{zhang12}, as well as our own single crystal data (see Supplementary Material).

Fig.~\ref{fig:evolution}(c) and (d) demonstrate the neutron scattering intensity of the cubic (111) peak as a function of temperature. Contrary to what is expected from structural considerations alone, NPD reveals a significant increase in this peak below the HTT$\rightarrow$FCO transition and again below the FCO$\rightarrow$LTT transition. At both temperatures, an increase in net moment has been noted in bulk magnetization data\cite{katsufuji08,zhang12}. For comparison, we also show in Fig.~\ref{fig:evolution}(d) the temperature evolution of the (220) peak from NPD and the (much weaker) (200) peak from single-crystal elastic scattering. The (220) peak carries only information about the iron sublattice, and its temperature dependence reflects only the 110 K transition. Conversely, the intensity of the (200) peak is zero at high-temperature and grows below the transition at 60 K. These observations are consistent with our Rietveld refinements, which associate two lowest phase transitions with the onset of collinear and canted ferrimagnetism, as discussed in more detail below.

\begin{figure}[t]
\begin{center}
\includegraphics[width=\columnwidth]{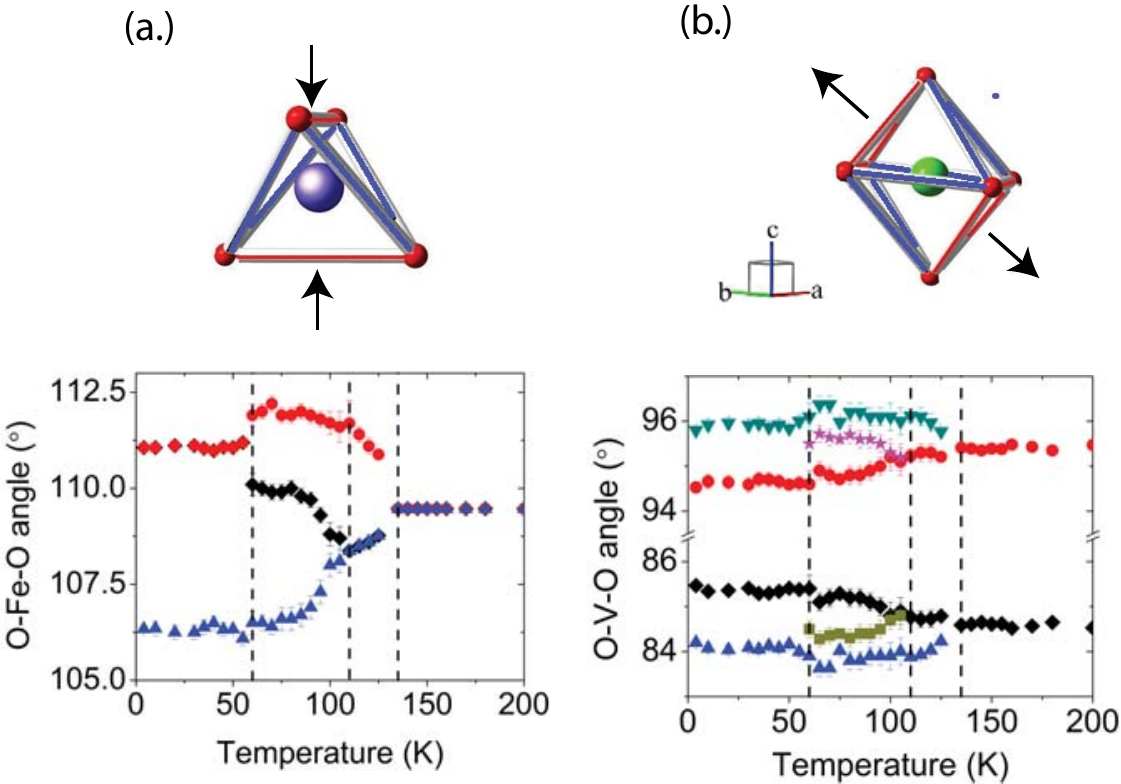}
\caption{Plots versus temperature of the interior $O-M-O$ angles ($M \in (Fe^{2+},V^{3+})$) of the (a) $FeO_{4}$ tetrahedra and (b) $VO_{6}$ octahedra, shown schematically at the top. In pictures, bonds of equal length are shown with equivalent coloring. Arrows denote the largest distortion of each polyhedron: a compression of the $FeO_{4}$ tetrahedron at 135 K along [001] and a trigonal distortion of the $VO_{6}$ octahedron along $<$111$>$ directions that exists at all temperatures. }\label{fig:polyhedra}
\end{center}
\end{figure}

To gain insight about the crystal field environment at the $Fe^{2+}$ and $V^{3+}$ sites, we further plot in Fig.~\ref{fig:polyhedra} the $O$-cation-$O$ angles as a function of temperature for the $FeO_{4}$ tetrahedra and $VO_{6}$ octahedra, respectively. In the high-temperature cubic ($Fd\bar{3}m$) phase, the interior angles of the $FeO_{4}$ tetrahedra (Fig.~\ref{fig:polyhedra}(a)) are sixfold-degenerate with a value of 109.4$^\circ$. This reflects the fact that the tetrahedra are regular, and thus the orbital ground-state of $Fe^{2+}$ is two-fold degenerate. At the 135 K transition, this degeneracy is lifted by a uniform compression in the cubic (001) direction, lessening four of the six $O-Fe-O$ angles and lowering the energy of the single-electron $d_{z^{2}}$ orbital relative to the $d_{x^{2}-y^{2}}$ orbital. As noted previously by Katsufuji \textit{et al.}\cite{katsufuji08} and Sarkar\textit {et al.}\cite{sarkar11}, this behavior can be associated with ferro-orbital order on the $Fe^{2+}$ site. The orthorhombic transition introduces an additional in-plane compression of the tetrahedra, which increases with decreasing temperature until the system locks-in to the low-temperature tetragonal structure. Notably, these subsequent transitions affect all $Fe^{2+}$ equivalently and are consistent with ferro-orbital order persisting to lowest temperatures (though the exact distribution of electron density will evolve upon cooling).

Conversely, in the $Fd\bar{3}m$ phase the $VO_6$ octahedra display a significant trigonal distortion from cubic symmetry. This is evident in Fig.~\ref{fig:polyhedra}(b), where one can see that at high temperatures the twelve interior angles of the $VO_{6}$ octahedra are split into two inequivalent branches of six with values significantly different from 90$^\circ$ (expected for regular octahedra). This splitting corresponds to a stretching of the octahedra in alternating $<$111$>$ directions along chains in the [110] direction. The distortion can be associated directly with the oxygen position parameter within the $Fd\bar{3}m$ space group. The angular separation at high temperatures (10.7$^\circ$) is comparable to what is inferred from the structures of $\mathrm{MnV_{2}O_{4}}$ ($12.9^\circ$) and $\mathrm{MgV_{2}O_{4}}$(8.7$^\circ$), where the importance of trigonal distortion has already been acknowledged\cite{sarkar09,chern10,wheeler10}. Katsufuji \textit{et al.} previously noted a tetragonal compression of the $VO_{6}$ octahedra at the 135 K transition\cite{katsufuji08}. This is observable in Fig.~\ref{fig:polyhedra}(b) as well. Notably however, the trigonal splitting of the O-V-O angles in the cubic state is nearly ten times larger than the perturbations at lower temperatures. (Note the break in the vertical axis.) Thus our data imply that the trigonal distortion is the dominant non-cubic term in the crystal field environment of $V^{3+}$.

\begin{figure}[t]
\begin{center}
\includegraphics[width=\columnwidth]{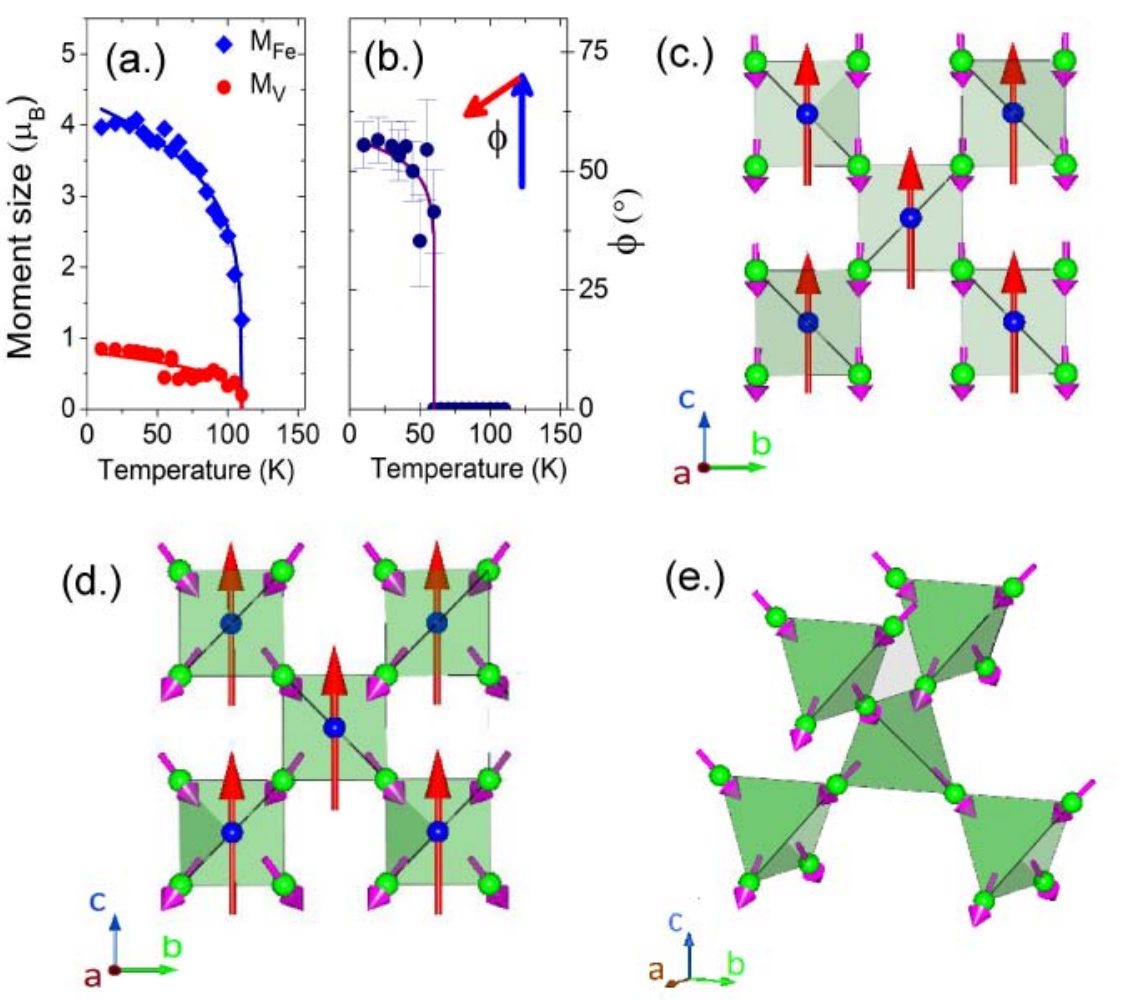}
\caption{\textbf{(a)} Plot of the ordered moment size on the $Fe^{2+}$ and $V^{3+}$ cation sites and \textbf{(b)} the angle between the $V^{3+}$ moments and the axis defined by $Fe^{2+}$ moments.  \textbf{(c)} A sketch of the collinear ferrimagnetic state in the FCO phase, as seen along the cubic (001) direction, and \textbf{(d)} a similar sketch of the canted state seen in the LTT phase. Details about the ordering patterns are given in the main text. \textbf{(e)} A view of $V^{3+}$ moments in low-temperature ordered state clearly demonstrating the 2-in-2-out spin structure on the pyrochlore sublattice.}\label{fig:magnetic}
\end{center}
\end{figure}

Information about the determined magnetic structure is displayed in Fig.~\ref{fig:magnetic}. As hypothesized previously in analogy to other spinel systems\cite{katsufuji08,zhang12}, the HTT$\rightarrow$FCO structural phase transition coincides with the onset of collinear ferrimagnetism, wherein $Fe^{2+}$ moments align ferromagnetically along the longest orthorhombic axis and $V^{3+}$ moments align antiparallel. This is the ordered state favored by antiferromagnetic superexchange between the $Fe^{2+}$ and $V^{3+}$ cations and implies that the transition is driven by such magnetic interactions. The orthorhombic structural transition then likely results from a coupling between cation spins and phonon modes which strongly affect superexchange paths, similar to what has been suggested by recent optical measurements of  $\mathrm{FeCr_{2}O_{4}}$ and $\mathrm{NiCr_{2}O_{4}}$\cite{bordacs09}. As shown in Fig.~\ref{fig:magnetic}(a), the ordered moment sizes on both sublattices grow below  $T_{N1}$ = 110 K, with the vanadium moment perhaps seeing additional increase at 60 K. The ordered moment sizes at lowest temperatures are 4.0$\mu_{B}$ and 0.85$\mu_{B}$ on the $Fe^{2+}$ and $V^{3+}$ sites, respectively. Whereas the number for $Fe^{2+}$ is in line with expectations for a S=2 system, the value for $V^{3+}$ is significantly less than the 2.0 $\mu_{B}$ expected for an S=1 cation with quenched orbital angular momentum.

Refinements indicate no tendency for $V^{3+}$ to cant away from the axis defined by the $Fe^{2+}$ moments in the orthorhombic phase, though symmetry would allow such behavior. Nonetheless, it is shown in Fig.~\ref{fig:magnetic}(b) that the refined angle between the $V^{3+}$ moment and the $Fe^{2+}$ spin axis jumps abruptly from zero to $\sim55^{\circ}\pm$4$^{\circ}$ below the FCO-LTT transition at $T_{N2}$ = 60 K. This canting transition is illustrated in Figs.~\ref{fig:magnetic}(c) and (d), where we sketch the ordered states above and below $T_{N2}$ as viewed along the cubic (100) direction. The existence of a canting transition is further supported by the single-crystal data in Fig.~\ref{fig:evolution}(d), as the intensity of the (200) Bragg peak is proportional to the square of the in-plane vanadium moment in this model. The low-temperature value of 55$^{\circ}$ for the canting differs from the 33$^\circ$ inferred recently by Kang \textit{et al.} from NMR data\cite{kang12}. However as noted previously\cite{kang12}, a value of 33$^\circ$ cannot be reconciled with the large ferrimagnetic moment observed via magnetization\cite{zhang12} (which implies an angle closer to 60$^\circ$). We feel that the neutron diffraction data provide the more reliable estimate.

The evolution from a collinear to canted ferrimagnetic state is reminiscent of the magnetic behavior of $\mathrm{MnV_{2}O_{4}}$, where the canting was accompanied by the gapping of magnetic excitations\cite{garlea08}. Significantly though, we note that the canting angle observed here is smaller than the 65$^\circ$ reported for $\mathrm{MnV_{2}O_{4}}$\cite{garlea08} and is equal within error to the 54.7$^{\circ}$ between the (001) and (111) directions in a cubic lattice. In addition to the canting angle, the `in-plane' (i.e. $\perp$(001)) spin directions imposed by the $I4{1}/amd$ space group are also different from what has been assumed for the ordered state of $\mathrm{MnV_{2}O_{4}}$\cite{garlea08}, and in fact refinement suggests the low-temperature phase of $\mathrm{FeV_{2}O_{4}}$ is characterized by a `2-in-2-out' structure on the pyrochlore lattice (see Fig.~\ref{fig:magnetic} (e)), more commonly associated with `ice-rules' in frustrated rare-earth antiferromagnets\cite{bramwell01_2}.

In $\mathrm{MnV_{2}O_{4}}$, the low temperature canting transition has been associated with the ordering of $V^{3+}$ orbital degrees-of-freedom\cite{garlea08,sarkar09,chern10}, and similar physics is likely at play here. Theoretically, it has been shown that the ground state properties of both $\mathrm{MnV_{2}O_{4}}$\cite{sarkar09} and $\mathrm{FeV_{2}O_{4}}$\cite{sarkar11} are strongly influenced by the alternating pattern of trigonal distortion axes which now has been observed in both materials. The resultant physics is captured by the `quantum 120$^{\circ}$ model' analysis of Chern \textit{et al}\cite{chern10}, who considered vanadium spinels with dominant trigonal distortions and showed for these materials that an orbitally ordered state is stabilized at low temperature with electron density modulated along $<$110$>$ directions. The spin order is then determined by a competition between relativistic spin-orbit coupling and orbital exchange between vanadium sites. When the latter wins out, the ordered spins tend to lie along primary cubic axes in the plane perpendicular to the ordered A-site spins. When spin-orbit coupling is dominant, a strong single-ion anisotropy forces the spins to lie along cubic $<$111$>$ directions and favors the `2-in-2-out' structure seen in Fig.~\ref{fig:magnetic}. It has been shown that $\mathrm{MnV_{2}O_{4}}$ is well-described by an intermediate case, where spin-orbit coupling and orbital exchange are comparable\cite{chern10}. The present data suggests that $\mathrm{FeV_{2}O_{4}}$ is well within the strong spin-orbit coupling limit of this model. It is worth noting that strong spin-orbit coupling would have the additional effect of reducing the $V^{3+}$  ordered moment size from spin-only values, consistent with observation.

From a pure physics perspective, the `2-in-2-out' spin structure itself may be one of the most interesting new observations. In the rare-earth pyrochlore antiferromagnets, a local `2-in-2-out' constraint arises from a net ferromagnetic coupling between Ising spins with quantization axis along $<$111$>$ and leads to the macroscopic ground-state degeneracy underlying the much-studied `spin-ice' class of compounds\cite{bramwell01_2}. Our data imply that in $\mathrm{FeV_{2}O_{4}}$ an equivalent Ising condition on vanadium spins results from the large trigonal distortion of the $VO_{6}$ octahedra in the presence of strong spin-orbit coupling. Thus, one way of interpreting the low-temperature magnetic state in this material is as a $d$-electron analogue of the spin-ice problem, but persisting to temperatures as high as 60 K. In such a picture, the role of the large ordered spin on the $Fe^{2+}$ site is to act as an applied field and break the macroscopic degeneracy; indeed, the magnetic order in the LTT phase is exactly what has been observed in spin-ice materials $Ho_2Ti_2O_7$ and $Dy_2Ti_2O_7$ with field applied along (001)\cite{fennel05}. An analogous situation may be found in the `ordered spin-ice' phases of $\mathrm{Sm_2Mo_2O_7}$\cite{singh08} and $\mathrm{Nd_2Mo_2O_7}$\cite{taguchi01}, where rare-earth moments on the pyrochlore lattice couple to ferromagnetically ordered $Mo$ spins. In $\mathrm{Tb_2Sn_2O_7}$, a similar `ordered spin-ice' state has also been seen\cite{mirebeau05}, and was initially associated with ferromagnetic near-neighbor coupling in the presence of moderate single-ion anisotropy. The results reported here provide an important link between such rare-earth pyrochlore antiferromagnets and trigonally-distorted spinels such as $\mathrm{FeV_{2}O_{4}}$.

This Research at Oak Ridge National Laboratory's High Flux Isotope Reactor was sponsored by the Scientific User Facilities Division, Office of Basic Energy Sciences, U. S. Department of Energy.  H.D.Z. is supported by NSF-DMR-0654118. The authors would additionally like to acknowledge Tao Hong for technical assistance during elastic neutron scattering measurements.

\clearpage
\setcounter{figure}{0}
\setcounter{equation}{0}
\setcounter{section}{0}

\begin{center}
\large
\textbf{SUPPORTING MATERIALS}
\end{center}

\section{Sample Preparation and Characterization}

The samples investigated in this manuscript were prepared by grinding appropriate mixtures of $\mathrm{Fe_2O_3}$, Fe and $\mathrm{V_2O_3}$, pressing under 400 atm hydrostatic pressure and calcining in flowing argon at 950 C for two days, with intermediate grinding to ensure sample homogeneity. The sample was characterized by bulk magnetization using a Quantum Design Magnetic Property Measurement System (MPMS), and the results are shown in Figure~S1. Of note are the distinct anomalies at the T = 110 K and 60 K, consistent with the ordered states inferred from our neutron scattering data.

A single crystal of $\mathrm{FeV_{2}O_{4}}$ was also recently been prepared via the float-zone method. The crystal growth was carried out in argon in an IR-heated image furnace equipped with two halogen lamps and double ellipsoidal mirrors.  Feed and seed rods rotated in opposite directions at 25 rpm during crystal growth at a rate of 25 mm/hr. In addition to magnetic susceptibility, a portion of the crystal was characterized via heat capacity using a Quantum Design Physical Property Measurement System (PPMS).
The results of these measurements are shown in (Fig.~S2) indicate second-order transitions with $T_c$ = 138, 107 and 60 K, consistent with our neutron powder diffraction results and heat capacity measurements on powders\cite{zhang12}.  Notably, there is no sign of a transition at 35 K, as reported by single crystal x-ray scattering studies\cite{katsufuji08}. This is discussed in more detail below.

\section{Neutron Scattering Measurements}

Powders were investigated using neutron powder diffraction, using the HB2A diffractometer at the High Flux Isotope Reactor\cite{HB2a}. High resolution data were collected at 200 K, 125 K, 75 K, 50 K and 4K using a 12$^{\prime}$- 21$^{\prime}$-6$^{\prime}$ collimation and two different wavelengths (1.538 \AA~ and 2.41 \AA). Additional lower resolution measurements were carried out using the wavelength 1.538 \AA~ and 12$^{\prime}$- 40$^{\prime}$-6$^{\prime}$ collimation to monitor the change in the scattering across various phase transitions. Rietveld refinements were performed on combined data sets from two separate wavelengths using FULLPROF\cite{fullprof}. Data over a limited range of 2$\theta$ and analysis are presented in the main text of this Letter. Diffraction patterns covering the entire accessible range of 2$\theta$ are shown in Figure~S3, and structural parameters inferred from Rietveld refinements are given in Table~\ref{table:structure_info}.

The single crystal was investigated using elastic neutron scattering. Measurements were performed using the cold neutron triple-axis spectrometer (CTAX) of HFIR with guide-open-80$^{\prime}$-open collimation. The energy of the scattered neutrons was fixed at $E_f$ = 5 meV. Higher order contamination was removed by a cooled Be filter placed between the sample and analyzer. Elastic scattering data pertinent to the current discussion were presented in Figure 2 of the main text. Further data will be presented in a future publication.

\section{Absence of a low-temperature orthorhombic phase}

Figure~S3 (and Figure~1 in the main text) shows high-resolution neutron scattering patterns for temperatures 200, 125, 75 and 4 K, chosen to be within each of the distinct phases identified by the x-ray scattering study of Katsufuji \textit{et al.}\cite{katsufuji08}. As mentioned in the main text, a high-resolution pattern also exists for T = 50 K, but was indistinguishable from the pattern at T = 4 K. This is consistent with Figure S2 above and Figure 2(a) of the main text, which show sign of a transition below the FTO$\rightarrow$LTT transition at 60K. This is in apparent contradiction, however, to the paper of Katsufuji \textit{et al.}, who identified a fourth, LTT$\rightarrow$LTO structural phase transition in a single crystal with $T_c$ = 35 K\cite{katsufuji08}. With regard to this point, it is interesting to note no sign of a LTT$\rightarrow$LTO transition was observed in either the powder diffraction data of Katsufuji \textit{et al.}\cite{katsufuji08} or the heat capacity data of Zhang \textit{et al.}\cite{zhang12}. We further note that absence of a peak at 35 K in our single crystal heat capacity data shown in Figure~S2. This last observation and the magnitude of the peak splitting reported in Ref~\onlinecite{katsufuji08} seem to preclude the idea that the LTT$\rightarrow$LTO transition is merely a subtle effect which is detectable only in single crystal samples.

An alternative explanation is that small differences in the composition of different samples could be responsible for the discrepancy in low-temperature behavior. It has been noted on several occasions that samples of $\mathrm{FeV_2O_4}$ display a strong tendency towards excess iron on the vanadium sites\cite{katsufuji08,zhang12,rogers63} (an effect also seen by us), and detailed study by Rogers \textit{et al.} has demonstrated that even small variations in iron stoichiometry can significantly affect the physical properties of this system\cite{rogers63}. This includes a change in the number of phase transitions observable with x-ray scattering above nitrogen temperatures (the limit of that particular study). Though a similar study has not been done at low temperatures to our knowledge, this material-specific sensitivity to iron concentration makes sample stoichiometry a prime candidate to explain the variation in the presence of the LTT$\rightarrow$LTO transition, even within Ref.~\onlinecite{katsufuji08}. We point to the variation in FCO-LTT transition temperatures between the powder ($T_c$ = 60 K) and crystal ($T_c$ = 85 K) samples of Ref.~\onlinecite{katsufuji08} as further support for this idea.

\section{Glass-like behavior in the FTO phase}

We also noted briefly in the main text that a spin-glass transition has been identified by Nishihara \textit{et al.} at 85 K using AC-susceptibility\cite{nishihara10}. Glassy correlations would appear in neutron measurements as diffuse scattering intensity centered around the locations of specific Bragg peaks. In our powder measurements of $\mathrm{FeV_2O_4}$, we did observe such scattering about the (111) position, however closer inspection reveals that the diffuse scattering intensity peaks at the 110 K tetragonal-orthorhombic phase transition, decreasing at lower temperatures with no further signature at 85 K. This is shown below in Figure~S4.

The more obvious interpetation of our diffuse scattering then is as critical correlations associated with a ferrimagnetic phase transition in a frustrated system, and not as a signature of glassiness. We do note however that these observations do not preclude the possibility of weak glassy dynamics associated with a different reciprocal lattice position, such as (200), but the intensity of the (111) scattering here makes powder diffraction an unsuitable vehicle for exploring such physics. In fact, the single crystal data presented in Figure 2(d) of the main text reveals significant correlated paramagnetic behavior above the 60 K transition which might be associated with the `glassy' dynamics of Ref.~\onlinecite{nishihara10}.

%\bibliography{./spinels_etal}

\begin{figure*}[h]
\begin{center}
\includegraphics[width=\columnwidth]{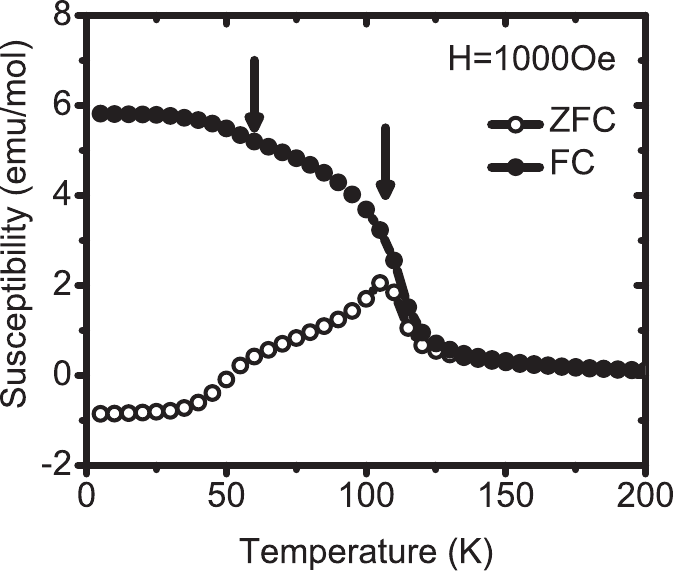}
\caption{Zero-field cooled (ZFC) and field-cooled (FC) measurements of bulk magnetization on powders of $\mathrm{FeV_2O_4}$. Details are in text of Supplementary Information.}\label{fig:squid}
\end{center}
\end{figure*}

\begin{figure*}[h]
\begin{center}
\includegraphics[width=\columnwidth]{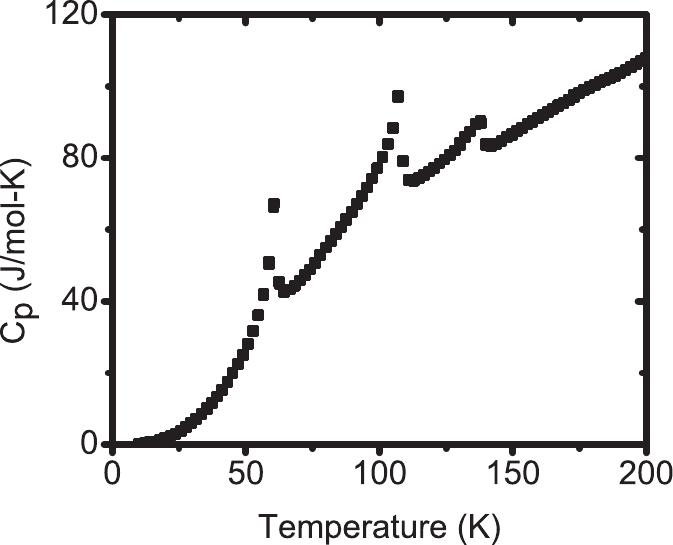}
\caption{Heat capacity of single crystalline $\mathrm{FeV_2O_4}$, as described in Supplementary Information text. Peaks imply transitions at 138 K, 107 K and 60 K, consistent with our results in the main text.}\label{fig:cp}
\end{center}
\end{figure*}

\begin{figure*}[h]
\begin{center}
\includegraphics[]{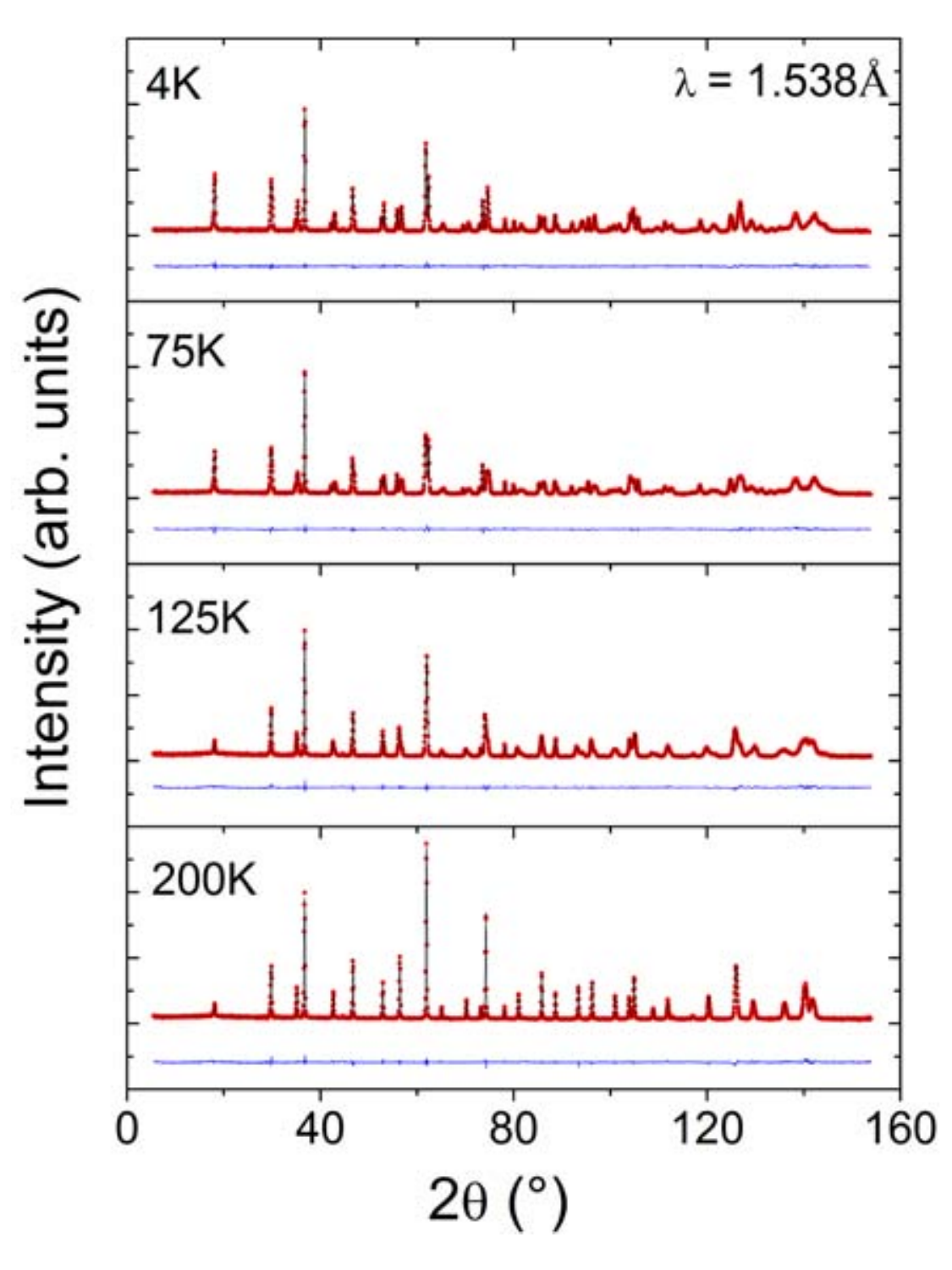}
\caption{Plots of raw NPD data (circles) measured at T = 200 K, 125 K, 75 K and 4 K. Solid lines are results of Rietveld refinements described in the main text. Differences between observed and calculated intensities are shown directly below the respective patterns.}\label{fig:raw}
\end{center}
\end{figure*}

\begin{table*}[h]
\small

\scalebox{0.9}{
\begin{tabular}{|c|c|c|c|c|}
	\hline
Temperature & 200 K & 125 K &  75 K & 4 K\\
	\hline
\centering

space group & $Fd\bar{3}m$  & $I4_{1}/amd$  & $Fddd$  & $I4_{1}/amd$\\
& (\#227) & (\#141) & (\#70) & (\#141)\\
a & 8.45547(5) & 5.99257(9) & 8.42930(15) & 5.94062(6)\\
b & 8.45547(5) & 5.99257(9) & 8.37082(14) & 5.94062(6)\\
c & 8.45547(5) & 8.40930(17) & 8.54962(11) & 8.54338(10)\\
%$Volume_{cubic}$ & 604.454(2) & 603.78(5)& 603.20(1) & 602.93(4)\\
\hline
$x_{Fe}$ & 0.375 & 0  & 0.125 & 0 \\
$y_{Fe}$ & 0.375 & 0.75 & 0.125 & 0.75\\
$z_{Fe}$ & 0.375 & 0.125 & 0.125 & 0.125\\
$B_{Fe}$ & 0.489(34) & 0.349(29) & 0.293(28) & 0.120(28)\\
\hline
$x_V$ & 0 & 0  & 0.5 & 0 \\
$y_V$ & 0 & 0 & 0.5 & 0\\
$z_V$ & 0 & 0.5 & 0.5 & 0.5\\
%$B_V$ & 0.5(0) & 0.5(0) & 0.20(0) & 0.20(0)\\
\hline
$x_O$ & 0.23888(7) & 0  & 0.26020(17) & 0 \\
$y_O$ & 0.23888(7) & 0.02367(23) & 0.25854(17) & 0.01820(19)\\
$z_O$ & 0.23888(7) & 0.25920(18) & 0.26419(12) & 0.26462(13)\\
$B_O$ & 0.461(25) & 0.428(22) & 0.320(23) & 0.243(23)\\

    \hline
\end{tabular}
}
\caption{Structural parameters extracted from simultaneous Rietveld refinement of $\lambda = 1.538 \AA$ and $\lambda = 2.41 \AA$  NPD data.}\label{table:structure_info}
\end{table*}

\newpage

\begin{figure*}[h]
\begin{center}
\includegraphics[width=\columnwidth]{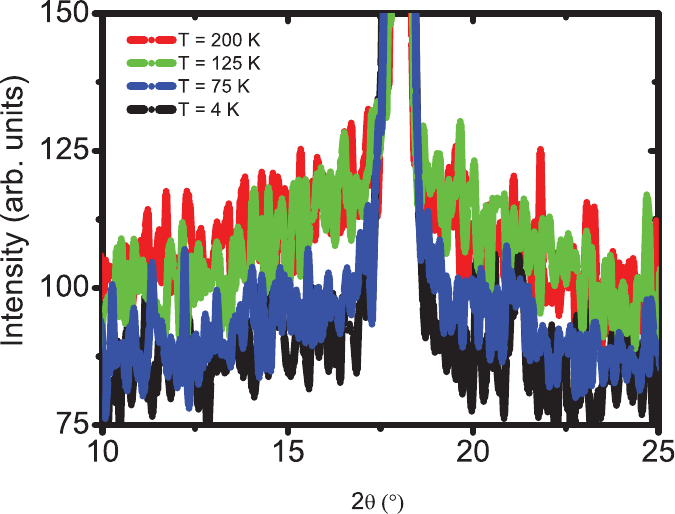}
\includegraphics[width=\columnwidth]{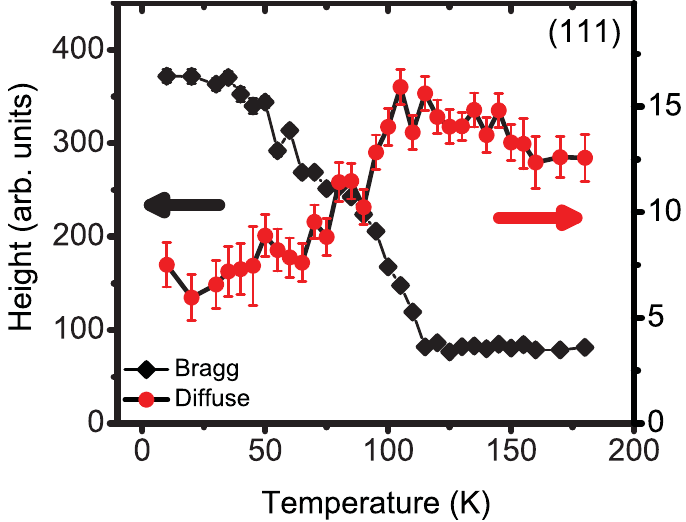}
\caption{(left) A plot of the weak diffuse scattering intensity about the (111) Bragg position for each of the patterns shown in Figure~S3. (right) The fit intensity of the diffuse scattering, as compared to the (111) Bragg intensity around the 110 K ferrimagnetic transition. }\label{fig:diffuse}
\end{center}
\end{figure*}

\end{document}